\documentclass{optica-article}

\journal{preprint} 

\articletype{Research Article}
\usepackage{eurosym}
\usepackage{lineno, siunitx}

\DeclareSIUnit{\sieuro}{\mbox{\euro}}

\begin{document}

\title{Pulsed laser diode excitation for transcranial photoacoustic imaging}

\author{Maxim N. Cherkashin,\authormark{\href{https://orcid.org/0000-0003-1283-7476} {\includegraphics[scale=0.07]{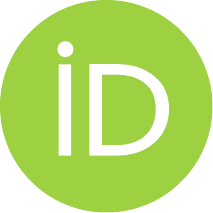}}, 1} Jan Laufer,\authormark{2} Thomas Kirchner\,\authormark{\href{https://orcid.org/0000-0002-3819-1987} {\includegraphics[scale=0.07]{orcid.pdf}},2,*}}

\address{\authormark{1} Photonics and Terahertz Technology, Faculty of Electrical Engineering and Information Technology, Ruhr University Bochum, Universitätsstrasse 150, 44780 Bochum, Germany
\\
\authormark{2} Medical Physics, Institut für Physik, Martin-Luther-Universität Halle-Wittenberg, Von-Danckelmann-Platz 3, 06120 Halle (Saale), Germany}

\email{\authormark{*} thomas.kirchner@physik.uni-halle.de} 

\begin{abstract*} 
Photoacoustic (PA) imaging of deep tissue tends to employ Q-switched lasers with high pulse energy to generate high optical fluence and therefore high PA signal. Compared to Q-switched lasers, pulsed laser diodes (PLDs) typically generate low pulse energy. In PA imaging applications with strong acoustic attenuation, such as through human skull bone, the broadband PA waves generated by nanoseconds laser pulses are significantly reduced in bandwidth during their propagation to a detector. As high-frequency PA signal components are not transmitted through skull, we propose to not generate them by increasing excitation pulse duration. Because PLDs are mainly limited in their peak power output, an increase in pulse duration linearly increases pulse energy and therefore PA signal amplitude.
Here we show that the optimal pulse duration for deep PA sensing through thick skull bone is far higher than in typical PA applications. Counterintuitively, this makes PLD excitation well-suited for transcranial photoacoustics. We show this in PA sensing experiments on \emph{ex vivo} human skull bone.
\end{abstract*}

\section{Introduction}

Photoacoustic (PA) waves are generated by pulsed light excitation of optical absorbers. This enables centimeter-deep acoustic imaging of optical absorption \cite{xu_photoacoustic_2006}, arising from chromophores, such as oxy- and deoxyhemoglobin. PA imaging of the brain is the subject of intense research \cite{yao_photoacoustic_2014}, as clinical applications of PA brain imaging promise functional assessment of perfusion and blood oxygenation that could assist in time-critical diagnosis of brain injury or stroke. To date, PA brain imaging has been demonstrated non-invasively in small animals \cite{wang_noninvasive_2003} and experiments using \emph{ex vivo} skull \cite{nie_photoacoustic_2011, kirchner_evaluation_2023}. \emph{In vivo} PA brain imaging in large animals \cite{kirchner_photoacoustics_2019} or humans \cite{na_massively_2021} has only been demonstrated following a craniectomy. 

For deep PA imaging, high excitation pulse energies are generally desirable to achieve a maximized signal-to-noise ratio (SNR). They are typically achieved using Q-switched excitation lasers. At fixed wavelengths, like \SI{1064}{nm}, Q-switched lasers can reach pulse energies up to $\sim$\SI{1}{J}, at a pulse repetition rate (PRR) of 1 to \SI{100}{Hz}. For a tunable near-infrared wavelength, optical parametric oscillators (OPO) are used, which provide pulse energies of tens of mJ. For \emph{in vivo} PA imaging, applicable PRRs and pulse energy are limited by the maximum permissible exposure (MPE) of skin. The pulse durations of Q-switched lasers are typically a few nanoseconds, in some cases tens of nanoseconds \cite{na_massively_2021}, which results in a broadband PA response, up to tens of MHz. 

As the acoustic attenuation of PA waves in skull bone is highly frequency dependent \cite{fry_acoustical_1978, liang_acoustic_2021, kirchner_evaluation_2023}, the transmitted acoustic power spectrum is cut off at 1-3\,MHz, depending on bone thickness and porosity. Therefore, the generated acoustic spectrum for transcranial PA imaging can be limited to a few MHz, which should allow for at least an order of magnitude longer PA excitation pulses.

Increasing pulse duration will not directly translate to an increase in pulse energy, and thereby SNR, when using typical Q-switched lasers. However, pulsed laser diode (PLD) arrays are mainly limited by the peak power output of the laser diodes and less so by pulse energy. In quasi-continuous wave (QCW) operation, the duty cycle of PLDs is limited to around 0.1\,\% to 1\,\%, allowing the selection of a wide range of pulse durations. Increasing PLD pulse duration can proportionally increase pulse energy at the cost of acoustic power at higher acoustic frequencies\cite{allen_generating_2005} -- but ideally only frequencies which will not penetrate skull bone anyway.

PLDs were first proposed as viable PA imaging excitation sources by Allen \emph{et al.} in 2005 \cite{allen_generating_2005, allen_pulsed_2006} and employed for \emph{in vivo} experiments shortly thereafter \cite{kolkman_vivo_2006}. During the last two decades, these sources have not seen widespread adoption in PA imaging \cite{upputuri_fast_2018}. However, QCW PLD stacks are well suited as a pump source for YAG lasers, which led to development of even higher power diodes, especially at \SI{808}{nm} and \SI{940}{nm} \cite{unger2021high}. Using these PLDs for PA imaging enabled highly integrated, potentially multi-wavelength illumination \cite{daoudi_handheld_2014}. These higher power LDs deliver pulse energies in the mJ range, leading to PA imaging research prototypes \cite{friedrich_quantitative_2011, daoudi_handheld_2014}. As compared in table \ref{tab:comparison}, currently available PLD stacks feature high PRRs \cite{sivasubramanian_high_2016}, a comparatively low cost and small footprint, and are generally used with longer pulse durations. Even though their pulse energies are significantly increased compared to earlier PLDs, they are still orders of magnitude lower than those of the highest-energy Q-switched laser sources.

\begin{table}[htbp]
\centering
\caption{ Comparison of typical current PA imaging excitation laser systems. }
\begin{tabular}{llll}
\hline
 & Q-switched & with OPO & PLD stacks\\
\hline
Pulse energy $E$ & $200$--$\SI{1000}{\milli\joule}$  & $10$--$\SI{50}{mJ}$  & $0.1$--$\SI{8}{mJ}$ \\
Maximum PRR & $1$--$\SI{100}{Hz}$&$10$--$\SI{100}{Hz}$&$200$--$\SI{6000}{Hz}$\\
Pulse duration $\tau$ &2--\SI{20}{ns} &2--\SI{20}{ns} &\SI{30}{ns} -- $\gg\SI{30}{ns}$\\
Wavelength & fixed
&tunable & fixed, multi-wavelength\\
Cost & $<\SI{100}{k}$€ & $>\SI{100}{k}$€& $\approx\SI{10}{k}$€\\ 
Footprint & moderate-high & high & low\\
\hline
\end{tabular}
\label{tab:comparison}
\end{table}

The combination of low pulse energy and long pulse durations produced by PLDs make them generally less useful for PA imaging than Q-switched lasers. However, specific PA imaging applications could benefit, especially applications with high PRR requirements, limitations in maximum permissible exposure (MPE) or high acoustic attenuation, leading to reduced acoustic bandwidth.

Acoustic attenuation is a limiting factor in achievable depth per resolution across PA imaging modalities \cite{wang_photoacoustic_2012}. In practice the detectable PA signal bandwidth is typically around 50 MHz for PA microscopy, and 5 MHz for PA tomography in centimeter depths \cite{wang_photoacoustic_2012, baumann2021backward}. The acoustic attenuation of human skull bone limits the bandwidth of transmitted photoacoustic signals to typically less than 2\,MHz \cite{kirchner_evaluation_2023}. Longer excitation pulse durations lead to PA waveforms of lower bandwidth which, in bandwidth-limiting imaging applications, will still lead to the same transmittable bandwidth -- without causing any loss in resolution. 

Increasing PLD pulse durations to match the generated acoustic frequency spectrum to the acoustic attenuation in soft tissue has been described theoretically\cite{allen_generating_2005, allen_pulsed_2006}. Due to even higher acoustic attenuation in skull bone than in soft tissues, transcranial PA allows for even longer pulses. PLDs can provide these longer pulse durations while also increasing PLD pulse energy and thereby generated PA signal amplitude. 

PLD emitters operate at a fixed wavelength, though multiple such emitters can be readily combined to allow for multispectral PA imaging, e.g.\ functional blood oxygenation measurements \cite{mienkina_multispectral_2010, friedrich_quantitative_2011}. This enables multispectral systems to be built at comparatively low cost and footprint. In contrast, available wavelengths for high power Q-switched lasers are limited, with \SI{1064}{nm} Nd:YAG lasers being the predominant PA imaging excitation laser. Lasers emitting at different wavelengths, e.g. alexandrite and ruby lasers, are rarely used and generally have lower power. Addition of harmonic generators and optical parametric oscillators (OPO) to Nd:YAG lasers offers a wide range of excitation wavelengths but at the cost of pulse energy; and also adds significant cost, complexity and footprint. 

In this work, we examine the effect of the duration of excitation pulses generated by different laser sources on PA signals transmitted through thick \emph{ex vivo} human skull bone. We demonstrate that the optimal excitation pulse duration for transcranial PA sensing is two orders of magnitude higher than typical PA excitation pulse durations. We discuss the suitability of PLDs as excitation sources for transcranial PA imaging and sensing.

\section{Materials and Methods}

To study the effect of excitation pulse duration and pulse shape on transcranial PA signals, the following three excitation lasers were used: 
\begin{itemize}
    \item[(1)] a Q-switched laser typical for PA imaging, 
    \item[(2)] a constant energy PLD system (CE-PLD), relatively short pulsed,
    \item[(3)] a constant power PLD system (CP-PLD), custom-made for wider variation in pulse duration.
\end{itemize}

As excitation source (1) we used a Nd:YAG laser (Quanta-Ray PRO-270-50, Spectra-Physics Lasers, Santa Clara, USA) at \SI{1064}{nm}. Pulse durations were varied between 40 and \SI{250}{ns} by adjusting the pump flash lamp voltage. The pulses were attenuated to obtain pulse energies of around \SI{1}{mJ}.

\begin{figure}[b!]
\centering\includegraphics[]
{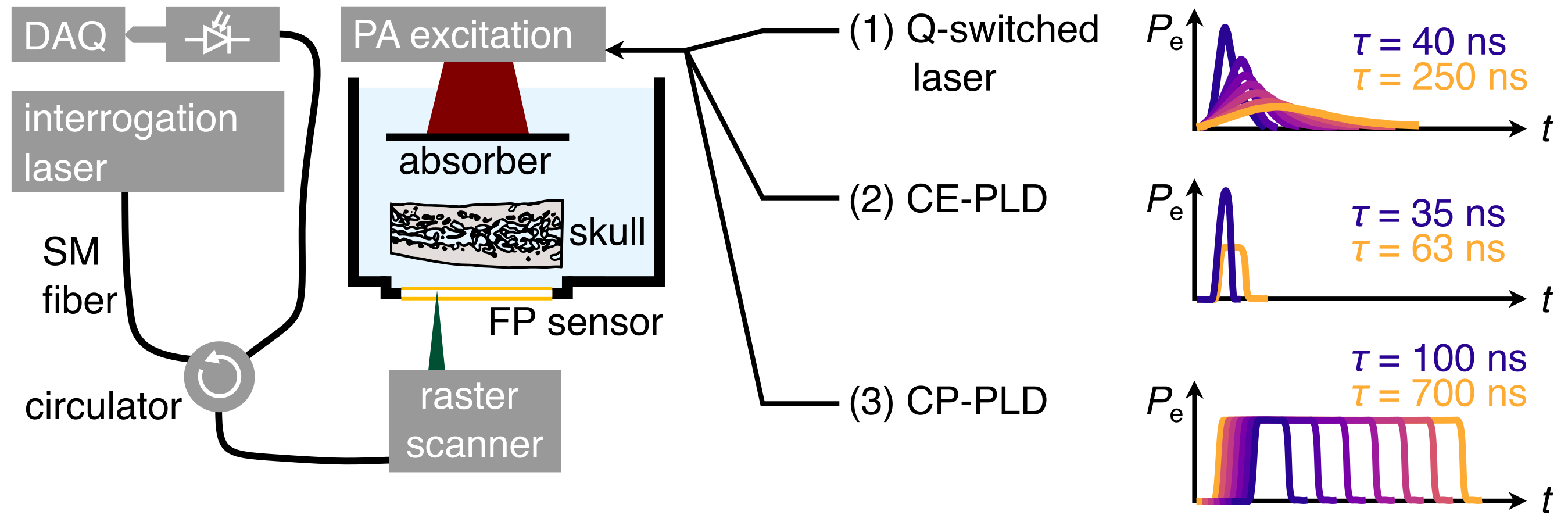}
\caption{Schematic of the experimental setup. A planar absorber is illuminated by one of three photoacoustic (PA) excitation sources: (1) a Q-switched laser, (2) a pulsed laser diode (PLD) system with constant energy (CE), (3) a PLD system with constant power (CP). The generated PA wave is measured using a planar FP sensor, interrogated by a PA tomography raster scanning setup similar to Zhang \emph{et al.}\cite{zhang_backward-mode_2008} -- using a narrow tunable laser as input and measuring the output voltage of a fast photodiode with a data acquisition (DAQ) card. An \emph{ex vivo} human frontal bone sample is placed in between the FP sensor and the PA source for acoustic transmission measurements. The shape and duration $\tau$ of the excitation pulse is varied as illustrated by the time-excitation-power curves $P_\textrm{e}(t)$ on the right.}
\label{fig:setup}
\end{figure}

The two PLD systems are based on vertical diode laser stacks emitting at a wavelength of \SI{808}{nm}. Both have a small size ($\sim$1\,L), compared to the Q-switched laser (>100\,L). The CE-PLD system is a commercial system (QD-Qxy10-IL, Lumibird, France) designed for PA imaging \cite{van_den_berg_feasibility_2017, upputuri_dynamic_2017} with a constant \SI{1}{mJ} pulse energy, up to \SI{2}{kHz} pulse repetition rate (PRR) and variable pulse duration between \SI{35}{ns} and \SI{63}{ns}. The CP-PLD system (PULS Direct Diode, Monocrom S.L., Barcelona, Spain) was custom-made and is operating at a maximum power of \SI{3.1}{kW}. It is designed to emit a wide range of pulse durations (\SI{100}{ns} to \SI{700}{ns}) yielding pulse energies between $0.3$ to \SI{2.1}{mJ} at PRRs of up to \SI{200}{Hz}. The pulse duration and PRR of the system are limited by the laser diode driver electronics and heat management. The output of the CP-PLD has a divergence angle of 10 degrees on the slow axis and 30 degrees on the fast axis. The output beam of the CE-PLD has a slow axis divergence angle of 10 degrees and a fast axis divergence collimated  to 3 degrees, using a microlens array. 

The experiments were conducted in transmission geometry in a water bath with sensor, skull and source placed along a common axis (see figure \ref{fig:setup}), similar to previous work \cite{kirchner_evaluation_2023}. To ensure broadband acoustic detection, a planar Fabry-Perot (FP) sensor with bandwidth of \SI{36}{MHz} and a flat frequency response \cite{buchmann_characterization_2017} is used. The PA response of the sensor is measured with a PA tomography raster scanner setup similar to Zhang \emph{et al.} \cite{zhang_backward-mode_2008,kirchner_evaluation_2023} using a CL band (\SI{1570}{nm}) interrogation laser (Tunics T100S, Yenista Optics, Lannion, France). PA waves were generated by illuminating a planar absorber (consisting of a thin film of black acrylic paint on an acrylic glass block) with the output of each excitation laser. The absorber was placed in a water bath, parallel to the FP sensor, at a distance of \SI{3}{cm} from the sensor. The pulse energy was measured using a beam sampler and a pyroelectric sensor (Ophir PE25-C, Ophir Optronics, North Andover, USA). Pulse duration and shape were measured with a fast silicon photodiode (UPD-500-SP, Alphalas, Göttingen, Germany).

In order to measure transcranial PA signals resulting from variable duration PLD PA excitation pulses, an \emph{ex vivo} frontal bone sample was inserted into the transmission path between PA source and FP sensor (figure \ref{fig:setup}). The skull sample \cite{kirchner_micro-ct_2022} consisted of frontal skull bone from a 70 year old male body donor. We chose this as a representative example of thick skull bone with strong acoustic attenuation \cite{kirchner_evaluation_2023}. Such \emph{thick} skull bone -- i.e.\ cranial bone consisting of Diploë sandwiched between cortical bone layers -- makes up the majority of the human skull. Written informed consent for general scientific investigation was given by the body donor.
To obtain reference measurements, PA signals transmitted through degassed water only (without skull) are recorded with each excitation laser. 

\section{Results}

We investigated the effect of pulse duration on the acoustic power density spectra of PA signals. Q-switched lasers are typically used for PA excitation with a fixed and short pulse durations. To generate close to Gaussian pulse shapes with variable pulse durations, we increased the pulse duration of a typical Q-switched Nd:YAG laser as described in section 2. All measurements are provided as Open Data (see 
Data Availability Statement).

\begin{figure}[b!]
\centering\includegraphics[width=5.25in]{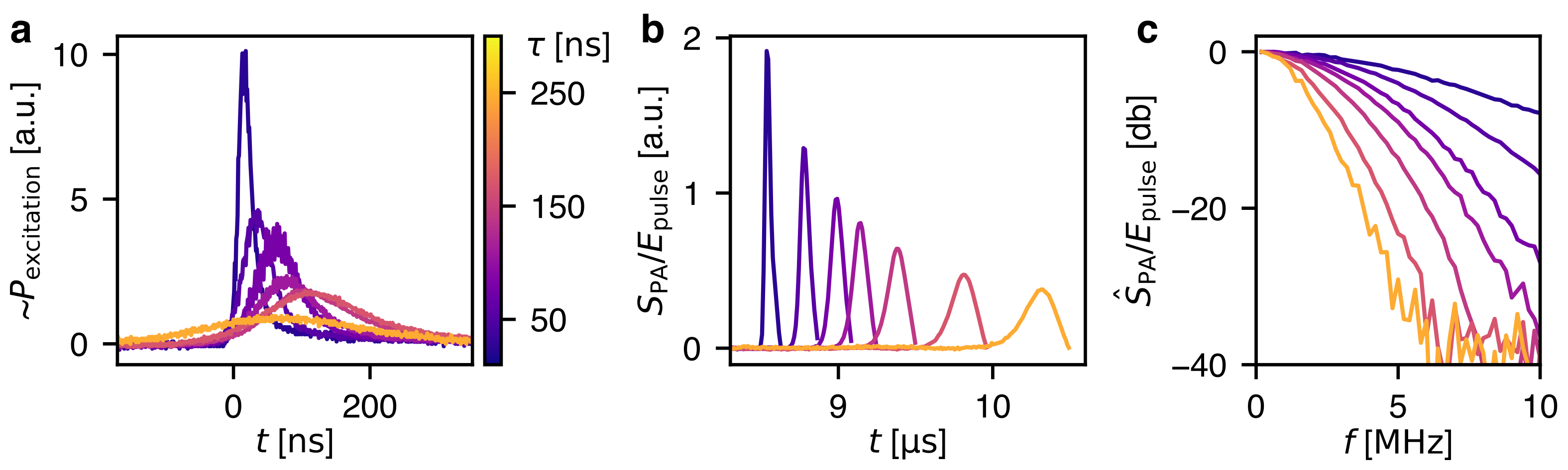}
\caption{PA excitation with a Q-switched laser with pulse durations $\tau$ of 40 to \SI{250}{ns}. \textbf{a} Pulse shapes as measured with a fast photodiode. \textbf{b} PA waveforms $S_\mathrm{PA}$ normalized for pulse energy $E$. \textbf{c} Acoustics power spectra -- Fourier transforms of the PA signals $\hat S_\mathrm{PA}$, normalized for pulse energy $E$.}
\label{fig:qswitched}
\end{figure}

Figure \ref{fig:qswitched}a shows the time-resolved excitation power for pulses of the Q-switched laser, measured from a reflection off a beam sampler. Figure \ref{fig:qswitched}b shows the PA signals detected by the ultrasound sensor after propagation through 1\,cm of water. The acoustic power spectra of the detected PA signals are shown in figure \ref{fig:qswitched}c. They are Fourier transforms of the energy-normalized signals in panel b. As expected, increasing the pulse duration $\tau$ decreases the bandwidth of the generated photoacoustic waveform. Figure \ref{fig:qswitched}c shows this behavior for Gaussian pulse shapes, which is in agreement with previous \emph{in silico} results \cite{allen_generating_2005}.

\begin{figure}[ht!]
\centering\includegraphics[width=5.25in]{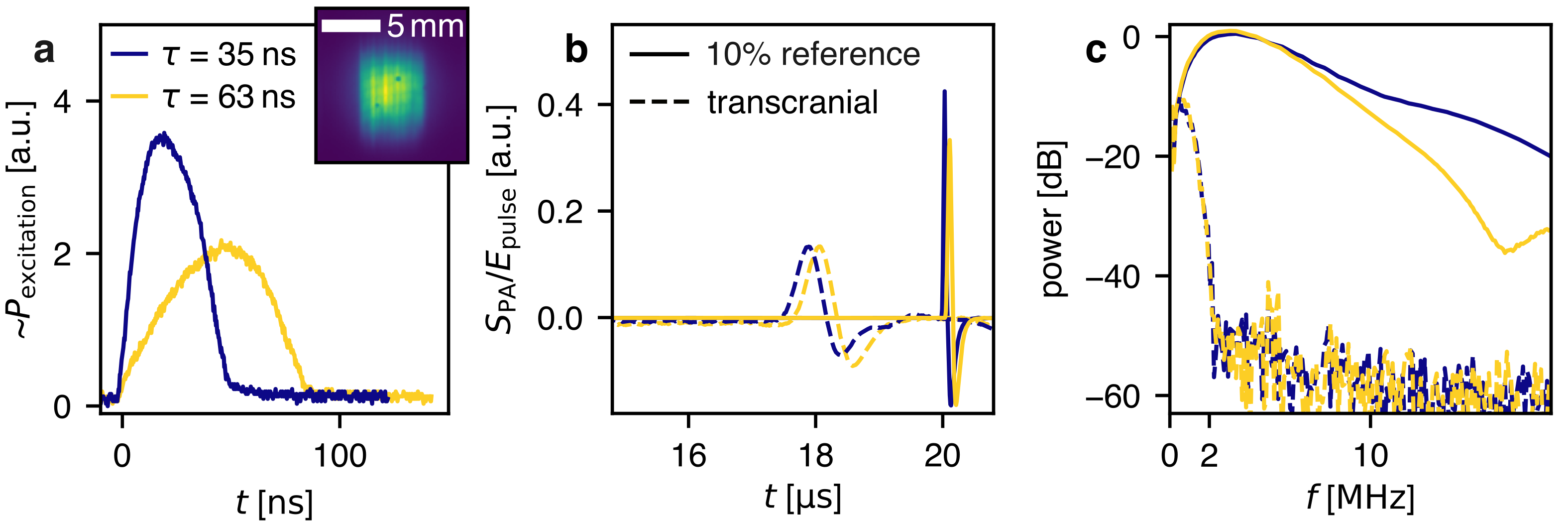}
\caption{Photoacoustic (PA) measurements of a thin, planar absorber using the constant energy pulsed laser diode (CE-PLD) system. \textbf{a} Pulse shape of the PA excitation pulses with pulse durations $\tau$. The reconstructed reference PA image is shown in the upper right corner as a maximum intensity projection along the depth axis. Single PLD bars collimated by microlenses can be distinguished. \textbf{b} Individual PA waveforms $S_\mathrm{PA}$ resulting from $\tau$ of \SI{35}{ns} and \SI{63}{ns}. Dashed lines are transcranial signals after passing through frontal bone. The solid line reference waveforms are shown at 10\,\% scale for visualization. \textbf{c} Acoustic power spectra for reference and transcranial signals.}
\label{fig:resultCE}
\end{figure}

Using the CE-PLD source we included transcranial PA sensing in the experiment.
PA waveforms were again excited in a planar thin film absorber with pulses shown in figure 3a. An insert in figure 3a shows the reconstructed initial pressure field in the absorber, illustrating the spatial distribution of the pulse energy on the planar absorber. The heterogeneity of the beam profile results from the collimation by a microlens array.
Figure \ref{fig:resultCE}b\&c shows PA signals generated using the longest (\SI{63}{ns}) and shortest (\SI{35}{ns}) pulse durations provided by the CE-PLD system. The signals are normalized with the pulse energy. The measured pulse energies were in fact similar (\SI{0.93}{mJ} and \SI{0.96}{mJ}). The PA waveforms shown in figure \ref{fig:resultCE}b were measured at the center of the FP sensor, on the acoustic axis. 1000 PA waveforms were averaged to increase SNR as the FP sensor had a relatively low sensitivity.

The PA measurement are performed in transmission with and without the skull sample. In the reference measurements (without skull), power spectra for varying excitation pulse durations differ only at frequencies above \SI{5}{MHz}, affecting the reference PA signal amplitude as seen in figure \ref{fig:resultCE}b. There is no significant difference between the transcranial PA signals, as virtually all acoustic power above \SI{2}{MHz} is attenuated in frontal bone. It is apparent that variations in pulse duration on such time scales (up to \SI{63}{ns}) have a negligible effect on transcranial PA signals.

\begin{figure}[ht!]
\centering\includegraphics[width=5.25in]{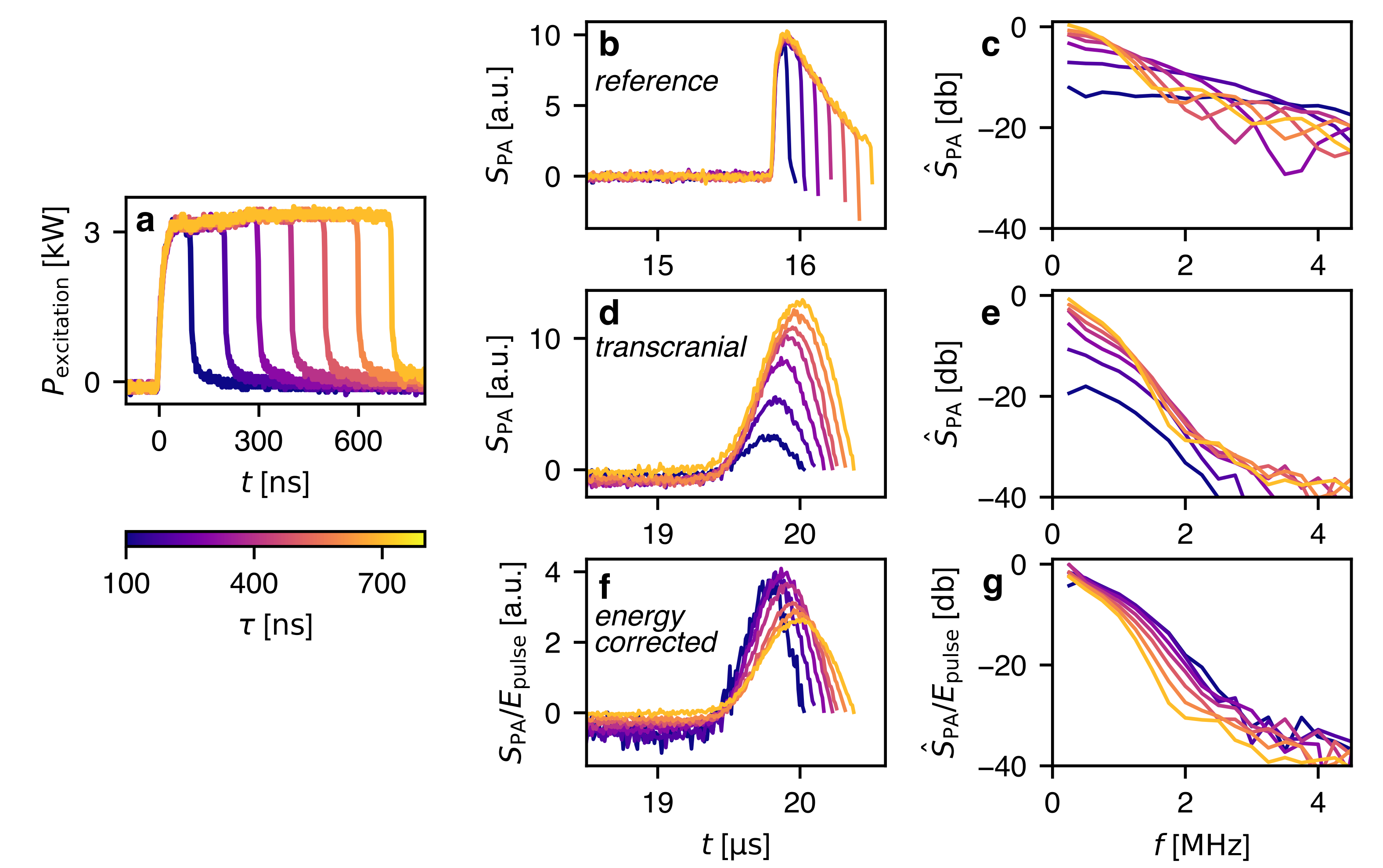}
\caption{PA transmission measurements through cranial bone using constant power pulsed laser diode (CP-PLD) excitation of varying excitation pulse duration $\tau$, ranging from $\SI{100}{ns}$ to $\SI{700}{ns}$ in increments of $\SI{100}{ns}$. \textbf{a} pulse shape measured using a fast photodiode. \textbf{b} Reference PA signals, and \textbf{c} their acoustic power spectra $\hat S_\mathrm{PA}$.  \textbf{d}~Transcranial PA signals, and \textbf{e} their acoustic power spectra $\hat S_\mathrm{PA}$. \textbf{f\&g} Pulse energy corrections of (d\&e).}
\label{fig:resultCP}
\end{figure}

To investigate longer timescales, we used the CP-PLD system with up to 700\,ns pulses. As shown in figure \ref{fig:resultCP}a, the CP-PLD system was driven at its maximum rated output power, producing square excitation pulses with nearly constant power. Rise and fall times of the square pulses were constant. 


Figure \ref{fig:resultCP}b shows PA waveforms detected after transmission through water. Their respective power spectra are shown in figure \ref{fig:resultCP}c. The transcranial PA waveforms, recorded after introducing the \emph{ex vivo} skull sample in the transmission path, are shown in figure \ref{fig:resultCP}d, with their respective power spectra in figure \ref{fig:resultCP}e. The pulse-energy-normalized transcranial PA waveforms and their power spectra are shown in figure \ref{fig:resultCP}f and figure \ref{fig:resultCP}g respectively. The PA waveforms shown in figure \ref{fig:resultCP} are averaged over $400$ pulses at \SI{200}{Hz} pulse repetition rate to increase SNR. 

With increasing pulse duration up to \SI{500}{ns}, the transmitted acoustic power (figure \ref{fig:resultCP}e) increases over the entire frequency range. For pulse durations longer than 500\,ns less acoustic power is generated at frequencies of \SI{2}{MHz} and below.

\section{Discussion}

The reference measurements, performed without skull bone, are consistent with previously described and theoretically expected behavior of long duration PA excitation pulses \cite{allen_generating_2005, allen_pulsed_2006}. Increasing the pulse duration of a square pulse (with constant rise/fall times and constant power, see figure \ref{fig:resultCP}a) does not generally lead to an increase of PA signal peak amplitudes but extends the duration of the PA waveform (see figure \ref{fig:resultCP}b) when observed with a sufficiently broadband detector. 

In transcranial PA measurements, skull bone acts as a higher-order low-pass filter for the PA waveform, while acoustic attenuation in soft tissue is far less frequency dependent. The transcranial PA signal amplitude increases with increased pulse duration of constant power (figure \ref{fig:resultCP}d), as the acoustic power transmitted through the frontal bone (figure \ref{fig:resultCP}e) increases over the full spectrum up to a pulse duration of $\tau = \SI{400}{ns}$, with $\tau \gtrapprox \SI{600}{ns}$ showing a slight additional attenuation around \SI{2}{MHz}. A pulse duration increase up to $\approx \SI{400}{ns}$ shows no reduction on PA signals transmitted through sufficiently thick skull bone, even when correcting for pulse energy (cf.\ figure \ref{fig:resultCP}g). Matching the pulse duration to the acoustic attenuation of the tissue does not by itself lead to an increase in SNR if the pulse energy is unchanged -- this can be seen in both figures \ref{fig:resultCE} and \ref{fig:resultCP}f. Because PLD sources are limited in their maximum power output, longer pulse durations increase the maximum pulse energy as well. Only these accompanying increases in the pulse energy from a PLD source lead to an increase in detectable, or transmitted, acoustic power.

The optimal pulse duration for the \emph{ex vivo} skull sample used in this study is approximately 400 to \SI{500}{ns}, without losses in any band of the acoustic spectrum. \emph{In vivo} skull is expected to cause similar acoustic attenuation. While skull thickness and therefore attenuation varies, the frontal bone sample used had a thickness which was representative of typical human skull bone surrounding the brain. In adult human skull generally, only the temporal bone is significantly more transparent to ultrasound -- such thin bone constitutes only a small part of the skull.

Considering excitation pulses on the order of hundreds of nanoseconds, current PLD stack systems can yield a PA imaging SNR similar to OPO excitation lasers, given the same acquisition duration. To give an example using typical lasers, we can compare a PLD with an OPO laser at the same wavelength (i.e. \SI{808}{nm}). PLD systems with pulse durations longer than \SI{100}{ns} can deliver \SI{5}{mJ} pulses with a PRR of \SI{3}{kHz}. OPO lasers can deliver pulse energies of approximately \SI{30}{mJ} at a pulse repetition rate (PRR) of \SI{100}{Hz}. Averaging 30 acquisitions will yield an acquisition rate of \SI{100}{Hz}, such a PLD source achieves an SNR equivalent to a typical OPO system at an order of magnitude lower cost, and at orders of magnitude smaller footprint in terms of scale and power consumption.
The reduced footprint makes PLD systems much more mobile and easy to integrate in medical devices. Pulse energies of PLD stacks are also stable in pulse energy over long acquisitions, with a typical standard deviation of less than 1\%. This would be useful for functional and quantitative PA imaging, where pulse energy variations are a source of estimation error.

A general limiting factor for imaging depth and SNR is the excitation laser fluence on the tissue surface, which is often limited by the maximum permissible exposure (MPE)\cite{ansi_american_2007}, defined in terms of pulse energy per skin surface area and dependent on pulse duration. Longer pulses allow for a safer energy delivery and lead to a higher MPE on skin. I.e.\ for pulse durations $\tau = \SI{1}{ns}$ to $\SI{100}{ns}$, MPE $\coloneq C_\mathrm{A}\cdot\SI{20}{mJ/cm\squared}$ (with $C_\mathrm{A} = 10^{2(\lambda[\si{\micro\meter}] - 0.7)}$ in the wavelength $\lambda$ range of \SI{700}{nm} to \SI{1050}{nm}) leading to an MPE of \SI{33}{mJ/cm\squared} (at $\lambda=\SI{808}{nm}$ and $\tau\leq\SI{100}{ns}$). For longer pulses with $\SI{100}{ns} < \tau < \SI{10}{s}$, MPE $ \coloneq C_\mathrm{A}\cdot\tau^{1/4}\cdot\SI{1.1}{J/cm\squared}$. Increasing the pulse duration to $\tau = \SI{500}{ns}$ leads to an increase in the $\textrm{MPE}$ to \SI{49}{mJ/cm\squared}. However, for laser exposure of more than 10\,s (up to $3 \cdot 10^5\si{s}$), the delivered average power is the relevant limit, with $\textrm{MPE} \coloneq C_\mathrm{A}\cdot\SI{0.2}{W\per cm\squared}$, $\textrm{MPE}(\SI{808}{nm}) = \SI{340}{mW\per\centi\meter\squared}$. 
This effectively limits the maximum PRR.

In addition, PA detection with high sensitivity for very low-frequency ultrasound is needed to make use of long-pulsed PLD excitation, as these low-frequency acoustic components in the PA signal increase linearly with PLD pulse duration. Low-frequency sensitive PA sensors would result in a bandwidth-matched detection to take full advantage of PLD excitation.

We can conclude that excitation pulses with durations of hundreds of nanoseconds increase the SNR of PA signals transmitted through thick cranial bone, provided the increase in pulse duration is accompanied by an increase pulse energies -- as is the case when using PLD excitation.

\begin{backmatter}
\bmsection{Funding}
This work was funded by the German Research Foundation (DFG, Deutsche Forschungsgemeinschaft) under grant numbers 471755457 and 499150341.

\bmsection{Acknowledgment}
The authors would like to thank Martin R. Hofmann for support and the provided equipment which was partially funded via ERC-FP7 Marie-Curie Actions (317526 OILTEBIA). The authors would also like to thank Heike Kielstein, Institute of Anatomy and Cell Biology, MLU Halle-Wittenberg for the provision of the human skull sample.

\bmsection{CRediT authorship contribution statement}
Maxim N. Cherkashin: Methodology, Validation, Investigation, Resources, Writing - Original Draft, Writing - Review \& Editing. Jan Laufer: Methodology, Resources, Writing - Review \& Editing. Thomas Kirchner: Conceptualization, Validation, Formal analysis, Investigation, Resources, Data Curation, Writing - Original Draft, Writing - Review \& Editing, Visualization, Project administration, Funding acquisition.

\bmsection{Disclosures}
The authors declare no conflicts of interest.

\bmsection{Data Availability Statement}
Data underlying the results presented in this paper are Open Data at \href{https://doi.org/10.5281/zenodo.14478306}{doi:10.5281/zenodo.14478306}


\end{backmatter}

\bibliography{bibl}
\end{document}